\documentclass{ws-procs975x65}
\newcommand{\be}{\begin{equation}}
\newcommand{\ee}{\end{equation}}
\newcommand{\bea}{\begin{eqnarray}}
\newcommand{\eea}{\end{eqnarray}}
\usepackage[spanish,english]{babel}
\usepackage[latin1]{inputenc}

\begin{document}

\title{{\bf Static Spherically Symmetric Solutions in Extended Palatini Gravity }}

\author{Gonzalo J. Olmo}

\address{Instituto de Estructura de la Materia, CSIC, Serrano 121, 28006 Madrid, Spain}

\author{Hèlios Sanchis-Alepuz}
\address{ Fachbereich Theoretische Physik, Institut für Physik, Karl-Franzens-Universität Graz,
Universitätsplatz 5, A-8010 Graz, Austria}

\author{Swapnil Tripathi}
\address{Physics Department, University of Wisconsin-Barron County, 1800 College Drive, Rice Lake, Wisconsin 54868, USA}

\begin{abstract}
We consider static spherically symmetric stellar configurations in 
Palatini theories of gravity in which the Lagrangian is an unspecified 
function of the form $f(R,R_{\mu\nu}R^{\mu\nu})$. We obtain the 
Tolman-Oppenheimer-Volkov equations corresponding to this class of theories 
and show that they recover those of $f(R)$ theories and General Relativity 
in the appropriate limits. We compute exterior vacuum solutions and comment on the possible expected modifications, as compared to GR, of the interior solutions. 
\end{abstract}

\bodymatter
\vspace{1cm}

The literature on modified theories of gravity in Palatini formalism, in which metric and connection are treated as independent fields, has been mainly restricted to Lagrangians of the form $f(R)$, where $R$ is the Ricci scalar, and their relation with the late time cosmic speedup. Palatini theories with Ricci squared terms have received much less attention due to some technical difficulties associated with the connection equation. The cosmology of $f(R_{\mu\nu}R^{\mu\nu})$ theories has been considered in some detail\cite{ABF04}, and it was found that the scalar $R_{\mu\nu}R^{\mu\nu}$ can be expressed as a function of the trace $T$ of the matter energy-momentum tensor, similarly as for the scalar $R$ in $f(R)$ theories. Theories of the form $R+ f(R_{\mu\nu}R^{\mu\nu})$ rather than
simply $f(R_{\mu\nu}R^{\mu\nu})$ have also been considered\cite{LBM07}, and a $3+1$ decomposition of the metric was introduced in order to solve perturbatively the connection equation. More recently, we managed to solve the connection equation in a fully covariant approach and without resorting to perturbative techniques\cite{OSAT09,OSAT09MG} for any $f(R,R_{\mu\nu}R^{\mu\nu})$ Palatini Lagrangian. We thus have accomplished all the necessary preliminary steps to put this family of theories in suitable form to carry out applications in different scenarios. In this work we use these recent results to work out the equations that govern the stellar structure of spherically symmetric, static stars. \\ 
\indent The action that defines a Palatini $f(R,R_{\mu\nu}R^{\mu\nu})$ theory is as follows
\begin{equation}\label{eq:action}
S[g,\Gamma,\psi_m]=\frac{1}{2\kappa^2}\int d^4x \sqrt{-g}f(R,Q) +S_m[g,\psi_m]
\end{equation}
where $g_{\alpha\beta}$ represents the space-time metric, $\Gamma^\alpha_{\beta\gamma}$ is the connection (which is independent of the metric),  $\psi_m$ represents the matter fields, $R=g^{\mu\nu}R_{\mu\nu}$, and $Q=R_{\mu\nu}R^{\mu\nu}$. Using the results of our previous works\cite{OSAT09,OSAT09MG}, the physical metric $g_{\mu\nu}$ is related to an auxiliary metric $h_{\mu\nu}$ by the following expressions
\begin{equation}
h_{\mu\nu}=\Omega\left( g_{\mu\nu}-\frac{\Lambda_2}{\Lambda_1-\Lambda_2} u_\mu u_\nu \right) \ , \
h^{\mu\nu}=\frac{1}{\Omega}\left( g^{\mu\nu}+\frac{\Lambda_2}{\Lambda_1} u^\mu u^\nu \right)
\end{equation}
where $\Omega=\left[\Lambda_1(\Lambda_1-\Lambda_2)\right]^{1/2}$, $\Lambda_1= \sqrt{2f_Q}\lambda+\frac{f_R}{2}$, 
 $\Lambda_2= \sqrt{2f_Q}\sigma$, $\lambda= \frac{1}{8}\left[3\sqrt{2f_Q}\left(R+\frac{f_R}{f_Q}\right)\pm\sqrt{{2f_Q}\left(R+\frac{f_R}{f_Q}\right)^2-8\kappa^2(\rho+P)}\right]$, $\sigma=\lambda\pm\sqrt{\lambda^2-\kappa^2(\rho+P)}$, and we have used the short-hand notation $f_R\equiv \partial_R f$, and $f_Q\equiv \partial_Q f$.
In terms of the auxiliary metric $h_{\mu\nu}$, the field equations for the metric can be written as
\begin{equation}\label{eq:Tmn-pfh}
R_{\mu\nu}(h)=\frac{1}{\Lambda_1}\left[\frac{\left(f+2\kappa^2P\right)}{2\Omega}h_{\mu\nu}+\frac{\Lambda_1\kappa^2(\rho+P)}{\Lambda_1-\Lambda_2}u_{\mu}u_\nu\right]\equiv \tau_{\mu\nu} \ .
\end{equation}
After some lengthy algebra, the three equations that determine the stellar structure in these theories can be written as follows ($S\equiv\Omega\Lambda_1/(\Lambda_1-\Lambda_2)$)
\begin{eqnarray}\label{eq:psir-fin}
\left(\frac{\Omega_r}{\Omega}+\frac{2}{r}\right)\psi_r &=&\frac{1}{A}\left[ \tau_r^r-\frac{\Omega}{S}\tau^t_t\right]     
-\frac{1}{2}\frac{\Omega_r}{\Omega}\left(2\frac{\Omega_r}{\Omega}+\frac{S_r}{S}\right)-\frac{1}{r}\left(\frac{S_r}{S}-\frac{\Omega_r}{\Omega}\right)+\frac{\Omega_{rr}}{\Omega}
\\
\label{eq:Mr-fin}
\left(\frac{\Omega_r}{\Omega}+\frac{2}{r}\right)\frac{M_r}{r}&=&\frac{3\tau_r^r-\frac{\Omega}{S}\tau_t^t}{2}+A\left[\frac{\Omega_{rr}}{\Omega}+\frac{\Omega_{r}}{\Omega}\left(\frac{2r-3M}{r(r-2M)}-\frac{3}{4}\frac{\Omega_{r}}{\Omega}\right)\right]\\
P_r&=&-\frac{P^{(0)}_r}{[1-\alpha(r)]}\frac{2}{\left[1\pm\sqrt{1-\beta(r)P^{(0)}_r}\right]} \label{eq:pressure-fin}
\end{eqnarray}
where $A(r)=1-2M(r)/r$, and we have used the following short-hand notation
\begin{eqnarray}
P^{(0)}_r&=& \frac{(\rho+P)}{r(r-2M(r))}\left[M(r)-\left(\tau_r^r+\frac{\Omega}{S}\tau_t^t\right)\frac{r^3}{4}\right] \\
\alpha(r) &=& \frac{(\rho+P)}{2}\left(\frac{\Omega_P}{\Omega}+\frac{S_P}{S}\right)\\
\beta(r) &=& (2r)\frac{\Omega_P}{\Omega}\left[1-\frac{(\rho+P)}{2}\left(\frac{3}{2}\frac{\Omega_P}{\Omega}-\left\{\frac{\Omega_P}{\Omega}-\frac{S_P}{S}\right\}\right)\right]
\end{eqnarray}
It can be shown that these equations smoothly recover the corresponding equations of $f(R)$ theories in the appropriate limit (and they also correct some typos present in an earlier work\cite{Olmo08b} of one of us). This limit  corresponds to $f_Q\to 0$, which leads to $\lambda\to f_R/\sqrt{8f_Q}$, $\sigma\to 0$, $\Lambda_2\to 0$, $\Lambda_1\to f_R$, $S\to \Omega\to f_R$. To get this result we must take the minus sign in front  of the squareroot of $\lambda$ and $\sigma$. \\
Let us now discuss some solutions of the above equations. It is easy to see that the exterior solution of the above equations is of the Schwarzschild-de Sitter type. Since outside of the star $\rho$ and $P$ vanish, $\Omega$ and $S$ become constants, and the equations boil down to $\psi_r=0$ and $M_r=(f/4\Lambda_1)|_{vac}r^2$, where $(f/4\Lambda_1)|_{vac}$ is evaluated in vacuum and plays the role of an effective cosmological constant. \\
For the discussion of interior solutions, it is useful to have in mind the following family of quadratic models: $f(R,Q)=R+aR^2/R_P+Q/R_P$, for which $R$ turns out to go exactly like in GR, $R=-\kappa^2T$. The stellar structure of quadratic $f(R)$ models has been discussed recently\cite{Olmo08b,quadratic}. If we take $a=-1/2$, we find\cite{OSAT09,OSAT09MG} 
\begin{equation}\label{eq:Q-1/2}
Q=\frac{3R_P^2}{8}\left[1-\frac{2\kappa^2(\rho+P)}{R_P}+\frac{2\kappa^4(\rho-3P)^2}{3R_P^2}-\sqrt{1-\frac{4\kappa^2(\rho+P)}{R_P}}\right] \ .
\end{equation}
At low energies, this expression recovers the GR limit, $Q\approx \left(3 P^2+\rho ^2\right)+\frac{3 (P+\rho )^3}{2 R_P}+\frac{15 (P+\rho )^4}{4 {R_P}^2}+\ldots$, but at very high energies, positivity of the argument in the square root of (\ref{eq:Q-1/2}) implies that $\kappa^2(\rho+P)\leq {R_P}/{4}$, which clearly shows that the combination $\rho+P$ is bounded from above. \\
\indent The interior solutions of the above equations must be very similar to those of GR except at the innermost regions of extremely compact objects, where the modified dynamics should play some important role. Since the differential equations have the same degree as those of GR, we do not expect new solutions which may depend on free parameters, as it happens in other types of modified theories which introduce higher order equations. Rather, the solutions of our set of equations must represent deformations of those found in GR. The GR solutions should be recovered smoothly and in a unique way in the limit $R_P\to \infty$. The fact that this family of models leads to bounded density and pressure raises a natural question: can we find static solutions corresponding to objects denser than the black holes of GR? This question is pertinent because in GR there can not be static solutions with $r-2M(r)\leq 0$, since they unavoidably lead to gravitational collapse and the divergence of energy density and curvature scalars. The fact that $\rho$ and $P$ are bounded in this theory suggests that such static solutions could exist. Exploring this possibility will be the subject of future work.  

\indent {\bf Acknowledgments.} H.S-A. has been partially supported by the Austrian Science
Fund FWF under Project No. P20592-N16.  G.J.O. thanks MICINN for a Juan de la Cierva contract, the Spanish Ministry of Education and its program ``José Castillejo'' for funding a stay at the CGC of the University of Wisconsin-Milwaukee, and the Physics departments of the UW-Milwaukee and UW-Barron County for their hospitality during the elaboration of this work. G.J.O. has also been partially supported by grant FIS2008-06078-C03-02.

\end{document}